\makeatletter
\newcommand{\dontusepackage}[2][]{%
  \@namedef{ver@#2.sty}{9999/12/31}%
  \@namedef{opt@#2.sty}{#1}}
\makeatother
\dontusepackage{subfigure}

\documentclass[english,]{sigchi-ext}

\usepackage{lmodern}
\usepackage{amssymb,amsmath}
\usepackage{ifxetex,ifluatex}
\usepackage{fixltx2e} 
\usepackage{ragged2e} 
\ifnum 0\ifxetex 1\fi\ifluatex 1\fi=0 
  \usepackage[T1]{fontenc}
  \usepackage[utf8]{inputenc}
\else 
  \ifxetex
    \usepackage{mathspec}
    \usepackage{xltxtra,xunicode}
  \else
    \usepackage{fontspec}
  \fi
  \defaultfontfeatures{Mapping=tex-text,Scale=MatchLowercase}
  
\fi
\IfFileExists{upquote.sty}{\usepackage{upquote}}{}
\IfFileExists{microtype.sty}{%
\usepackage{microtype}
\UseMicrotypeSet[protrusion]{basicmath} 
}{}
\ifxetex
  \usepackage{polyglossia}
  \setmainlanguage{}
\else
  \usepackage[shorthands=off,english]{babel}
\fi
\usepackage{listings}
\usepackage{graphicx}
\makeatletter
\def\maxwidth{\ifdim\Gin@nat@width>\linewidth\linewidth\else\Gin@nat@width\fi}
\def\maxheight{\ifdim\Gin@nat@height>\textheight\textheight\else\Gin@nat@height\fi}
\makeatother
\setkeys{Gin}{width=\maxwidth,height=\maxheight,keepaspectratio}

\usepackage[font={small},skip=2pt]{caption}
\usepackage{float}
\hypersetup{breaklinks=true,
            bookmarks=true,
            pdfauthor={},
            pdftitle={Remote Heart Rate Sensing and Projection to Renew Traditional Board Games and Foster Social Interactions},
            colorlinks=true,
            citecolor=black,
            urlcolor=blue,
            linkcolor=black,
            pdfborder={0 0 0}}
\usepackage{balance}

\title{Remote Heart Rate Sensing and Projection to Renew Traditional Board
Games and Foster Social Interactions}

\author{
     \alignauthor{%
    \textbf{Jérémy Frey}\\
              \affaddr{Univ. Bordeaux} \\
        \affaddr{351 Cours de la Libération} \\
        \affaddr{33400 Talence, France} \\
         \email{jeremy.frey@inria.fr}\\
 } 
   }

\date{}

\usepackage{balance}

\usepackage{comment}

\usepackage[scaled=.92]{helvet} 

\usepackage{suffix}

\usepackage{marginnote}
\reversemarginpar

\copyrightinfo{Permission to make digital or hard copies of part or all of this work for personal or classroom use is granted without fee provided that copies are not made or distributed for profit or commercial advantage and that copies bear this notice and the full citation on the first page. Copyrights for third-party components of this work must be honored. For all other uses, contact the Owner/Author.\\
Copyright is held by the owner/author(s).\\
{\emph{CHI'16 Extended Abstracts}}, May 07-12, 2016, San Jose, CA, USA.\\
ACM 978-1-4503-4082-3/16/05. \\
http://dx.doi.org/10.1145/2851581.2892391}

\begin{document}


\maketitle

\RaggedRight{} 

\begin{abstract}
While physiological sensors enter the mass market and reach the general
public, they are still mainly employed to monitor health -- whether it
is for medical purpose or sports. We describe an application that uses
heart rate feedback as an incentive for social interactions. A
traditional board game has been ``augmented'' through remote
physiological sensing, using webcams. Projection helped to conceal the
technological aspects from users. We detail how players reacted --
stressful situations could emerge when users are deprived from their own
signals -- and we give directions for game designers to integrate
physiological sensors.
\end{abstract}

\keywords{
      Heart Rate;
      Social Presence;
      Board Games;
      Physiological Computing;
      Spatial Augmented Reality}

      \category{H.1.2}{User/Machine Systems}{Human information processing}
      \category{H.5.1}{Multimedia Information System}{Artificial, augmented, and virtual realities}

\def \citep {\protect\cite}

\WithSuffix\newcommand\caption*{\caption}

\reversemarginpar
\marginnote{
\begin{minipage}{\marginparwidth}
   \vspace{-50ex}
   \includegraphics[width=1\marginparwidth]{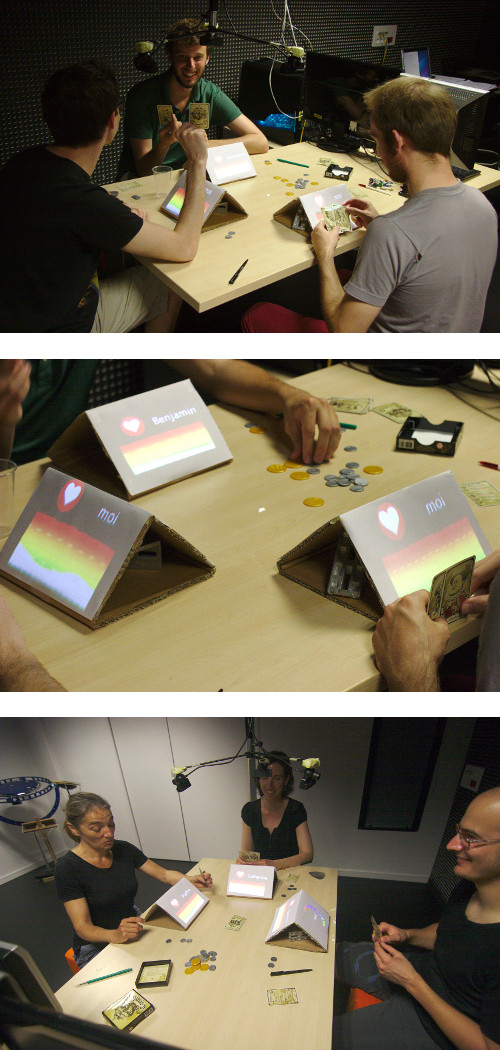}
   \captionof{figure}{A board game session that we augmented with remote physiological monitoring and projection.
    \label{fig:trailer}
   }
\end{minipage}
}

\section{Introduction}\label{introduction}

Through the rise of wearables -- such as smartwatches -- physiological
sensors are gaining increased attention. However, for the general
public, the range of use-cases of such sensors seems mostly limited to
health and performance. More often than not, physiological sensing is
experienced within medical or sports settings only. Heart rate belts,
for example, are sold in many sports shops across the globe to help
sportspeople boost their performance.

Physiological sensors can bring more to society. Lately, they have been
investigated as a supplementary communication channel. A study showed
that heartbeats were a meaningful source of information that could help
people to ``connect'' between each others \citep{Slovak2012}. In the
recent years, sensors have been used to mediate affect
\citep{Williams2015}, to support social interplay \citep{Walmink2014} or
to enhance telepresence \citep{Lee2014}.

Our work is part of the same movement, that tries to leverage social
interactions among peers with physiological computing. Indeed, such
information could help to create deeper interactions
\citep{Janssen2011}, enriching social presence -- that relates to the
degree of salience of another person \citep{Short1976}. We focus on
playful and casual interactions, because it is also an opportunity to
foster the \emph{use} of physiological sensors among the general public.
To do so, we propose to integrate heart rate sensing to a traditional
board game.

In the present work we used projection to display information. It is
less likely to disrupt the gaming experience than relying on screens
because the projected content can be integrated seamlessly to the
environment. Indeed, spatial augmented reality (SAR, introduced in
\citep{Raskar2001}) brings digital content to the physical world and
thus facilitates the merge between computers and existing board games
(Figure \ref{fig:trailer}).

Furthermore, as opposed to sensors attached to the body, we relied on a
non-contact system to record heart rate by the mean of
photoplethysmographiy (PPG). It uses webcams to process subtle
variations in skin's color while blood is flowing. To our knowledge,
this is the first time that remote sensing is used for several users, in
a social context.

\marginnote{
\begin{minipage}{\marginparwidth}
   \vspace{-80ex}
   \includegraphics[width=\marginparwidth]{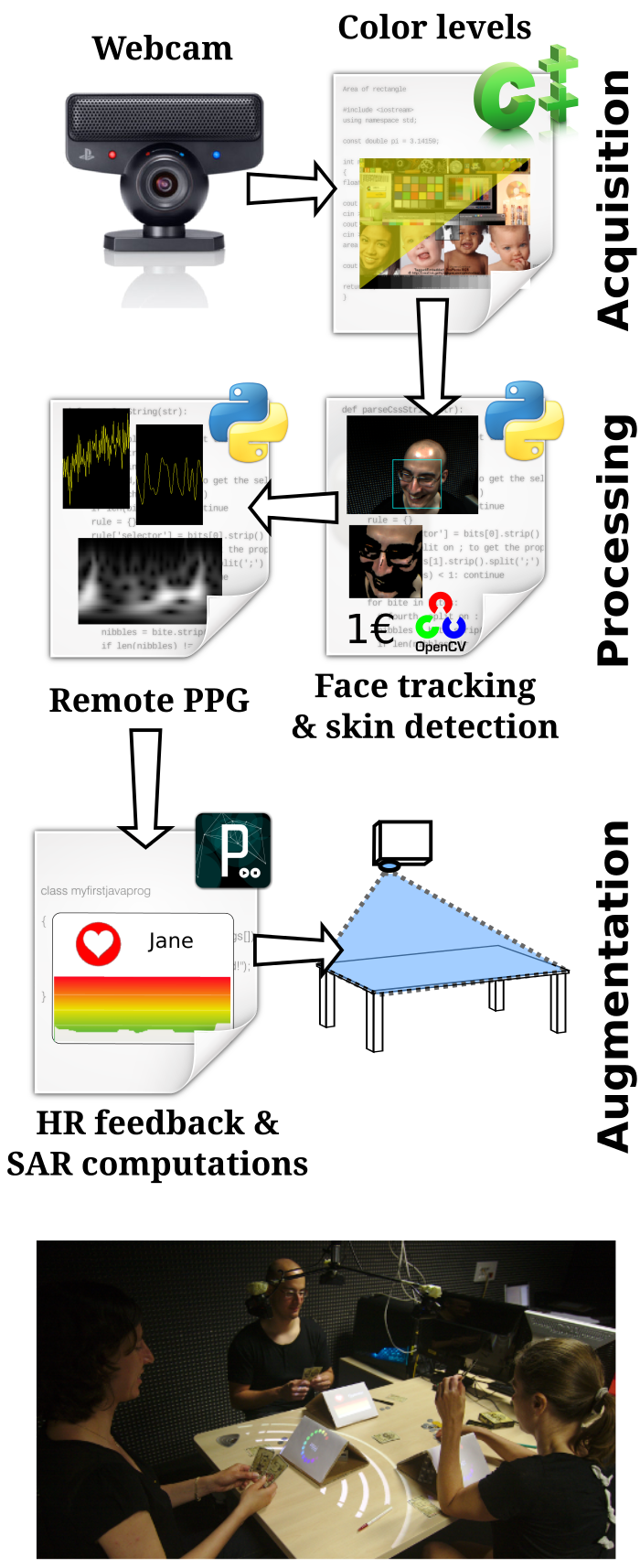}
   \captionof{figure}{The pipeline composing our system.
    \label{fig:pipeline}}
\end{minipage}
}

Using SAR and PPG, the technology disappears in the eyes of the players,
keeping the genuine ``feel'' of traditional board games. But then we
need to ensure that such addition does not hinder user experience and
that sharing an information that usually belongs to the realm of the
self does not deter how users feel. Some may not like that others see
``through'' them, especially when heartbeats may relate to intimacy
\citep{Janssen2011}. In particular, this negative effect may be more
likely to arise if the situation between players is not perceived as
being ``fair'', e.g.~if the biofeedback is seen by others and \emph{not
by the one being measured}.

Our first hypothesis is that the presence of a biofeedback equally
shared between players -- i.e.~a heart rate visible by all -- will
improve game experience and social presence. Our second hypothesis is
that an \emph{asymmetrical} biofeedback -- i.e.~players see opponents'
heart rate but not their own -- will on the contrary cause more
stressful situations and \emph{deter} game experience.

In order to test these hypotheses, we used the versatility offered by
SAR to create three different biofeedback conditions: heart rates
visible by all, heart rates visible by others but not by self, no heart
rates displayed.

The main highlights of this paper are:

\begin{enumerate}
\def\labelenumi{\arabic{enumi}.}
\itemsep1pt\parskip0pt\parsep0pt
\item
  To integrate seamlessly physiological computing into a traditional
  board game
\item
  To investigate how biofeedack influences user experience and social
  interactions
\end{enumerate}

\section{Related work}\label{related-work}

Even though previous works combined physiological monitoring and board
games, they did not focus on how users reacted to this new feedback --
nor did they consider the benefits for board games in general. For
instance, \citep{Slovak2012} investigated how people comprehend heart
rate feedback in various situations, and the appearance of a gaming
application among users was incidental. The biofeedback was studied for
training in \citep{Yamabe2010} as a way to help poker players gain
control over themselves. Another combination of poker and physiological
signals is sketched in \citep{Dang2010}, but heart activity only stands
as an additional feature of a new human-computer interaction technique.
Our interest, on the other hand, is from the start in human-\emph{human}
interactions.

We seek to use a game as a dedicated use case of physiological
monitoring's influence over social interactions. We also want to explore
how we could maximize user experience by integrating seamlessly the
technology behind. As a matter of fact, the tabletop setup proposed in
\citep{Dang2010} requires additional gestures from users to perform
actions as simple as hiding cards and in \citep{Slovak2012} each player
needed a laptop. Yamabe and al. \citep{Yamabe2010} used a projector to
display the heartbeat directly on the gaming table, but, as in each
other previous work, users still had to wear sensors. It is possible for
technology to be even less intrusive, and we solution both kinds of
artifacts.

\section{Description of the system}\label{description-of-the-system}

The main idea is to propose a ``sit and play'' setup for 3 persons,
where players would not have to endure any supplementary equipment
before experimenting physiological sensing, and with no computers inside
the game area in order to keep the genuine feel of board games.

Rather than using electrocardiography -- that requires electrodes on the
torso or on the wrists --, or other contact sensors, we turned to video
analysis to record heart activity. To embed the visual feedback in the
surroundings, we used projection instead of screens -- this design
choice also helped to present an asymmetrical biofeedback to players
(Figure \ref{fig:conditions}, middle).

While we cannot describe below all the technical aspects related to our
system due to space limitation, we release our entire pipeline --
summarized by Figure \ref{fig:pipeline} -- as an open-source software,
for others to benefit from the technology\footnote{\url{https://github.com/jfrey-xx/PhysioBluffGame.meta}}.

\subsection{Heart rate measures}\label{heart-rate-measures}

Subtle color changes in a video could be amplified to the point that the
variations of skin pigmentation occurring along each heartbeat become
visible. To get real-time measures from 3 persons at the same time, we
implemented an algorithm that takes values averaged over the face and
that is computationally efficient. It was presented by
\citep{Bousefsaf2013}, and enables a good accuracy even with regular
webcams.

The optical measuring of the volumetric variation of an organ -- such as
the heart -- is dubbed as photoplethysmography (PPG), hence this
video-based method is called \emph{remote} PPG. We successfully
integrated various webcams into our workflow. We validated our
implementation in a separate study by comparing remote PPG measures to a
ground truth obtained with an electrocardiogram (ECG). Over 10 minutes
recording sessions, the Pearson correlation varied between 0.30 and 0.81
-- Figure \ref{fig:ppg}, see also \citep{Frey2015c}.

\marginnote{
\begin{minipage}{\marginparwidth}
   \vspace{-40ex}
   \includegraphics[width=\marginparwidth]{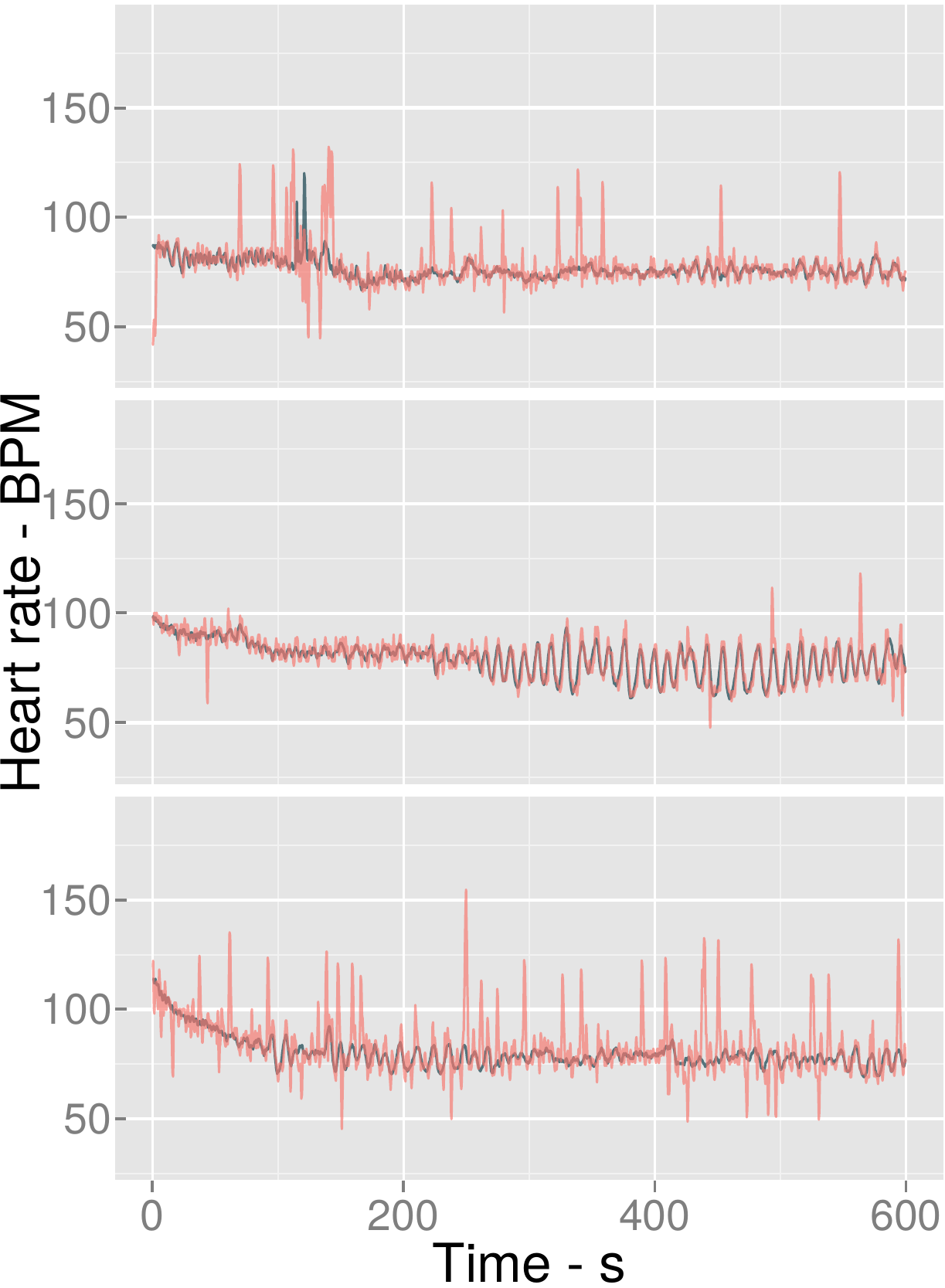}
   \captionof{figure}{Separate study to validate remote PPG: heart rate measures over 3 recording sessions of 10 minutes. Remote PPG is represented in red, the ground truth (ECG) in black. Average Pearson correlation: 0.53. Remote PPG is accurate enough to account for the heart rate variability caused by deep breathing -- e.g. oscillations in session 2, \emph{middle}. Each session was preceded by 5 minutes of aerobic exercise. See \citep{Frey2015c} for more details.
    \label{fig:ppg}}
\end{minipage}
}

\subsection{Spatial augmented reality}\label{spatial-augmented-reality}

Spatial augmented reality merges the digital world into the physical
world. It made the computers ``disappear'' in the eye of the players,
helping physiological monitoring to become part of a traditional board
game.

SAR was also a way to multiply the displays without the need of adding
physical screens -- tablets, laptops, \ldots{} -- to players'
surroundings. For instance, since we wanted to compare whether or not
the visual feedback of oneself heart rate would change game experience,
we just had to craft display stands with two sides, onto which we
projected either a heart rate or an idling animation (see Figure
\ref{fig:conditions}), instead of using 6 separate screens.

The video projector was positioned in a top-down orientation 1.5m above
the table. The resulting display surface was 1.2m by 0.75m. The visual
feedback of the heart rate had two modalities. An icon shaped as a heart
that was beating at the pace recorded by PPG, and beneath was a
histogram plotting the BPM (beats per minute) of the previous 20
seconds. The names of the players were displayed on the stands' sides
facing others -- ``me'' on the side facing them, helping to raise both
their presence and their social awareness. To obtain the desired
visualization we used a framework developed in Processing\footnote{\url{https://processing.org/}}
that could be easily grasped by game makers or
artists\citep{Laviole2012b}.

All computations, for all 3 players, were done on a single computer, an
Alienware Aurora R4 with an Intel i7-3820 processor, 8GB of RAM and a
GeForce GTX 660 Ti graphic card running Kubuntu 14.04 operating system.

\section{Pilot study}\label{pilot-study}

With the first iteration of our system, we wanted to investigate how
users felt regarding the heart rate feedback and the setup. Our
hypothesis is that while physiological computing enhance game
experience, more stressful situations could arise if the feedback is
asymmetric, not visible by self.

In this study we compared three different conditions of heart rate
feedback: heart rate visible by all (``HR all''), heart rate visible by
the others but not by self (``HR others''), heart rate visible by none
(control condition, ``HR none'') -- see Figure \ref{fig:conditions}. We
used a within-subject experimental design. The conditions were set for
all 3 players of a group at the same time, and each condition occurred
once. The order of the conditions was counter balanced between groups
following a latin square -- hence we recruited 6 groups.

We used a questionnaire inquiring about social presence to measure the
main effect of our experimental design and collated players' spontaneous
comments to gain further insight about the overall game experience.

\subsection{Board game}\label{board-game}

We chose a friendly and casual board game known as ``Coup'' -- edited by
\emph{Indie Boards and Cards}\footnote{\url{http://www.indieboardsandcards.com/}}
-- in its French version, ``Complots''. \emph{Coup} possesses bluffing
as one of the core elements of its gameplay. This is an incentive for
players to use the physiological signals, since for the general public
heart rate is strongly related to emotions. In \emph{Coup} each player
is given two random cards, each card representing a character, each
character having a ``power'' -- block or counter attacks, steal money,
and so on. The goal is to ``kill'' the characters of the other players.
There are various occasions to interact, and unless someone is
``challenged'' by an opponent, players never have to actually
\emph{show} their cards. These situations are most engaging for players.
The pace is fast, one game lasts about 5 minutes.

\subsection{Participants \& Apparatus}\label{participants-apparatus}

18 participants took part in this study -- 6 groups of 3 players, 5
females, 13 males, mean age 23.3 (SD: 6.9). Most of the participants
knew each others beforehand. Half of them reported a previous use of
physiological sensors, each time associated to sport or to medical
activities.

\marginnote{
\begin{minipage}{\marginparwidth}
   \vspace{-50ex}
   \includegraphics[width=0.9\marginparwidth]{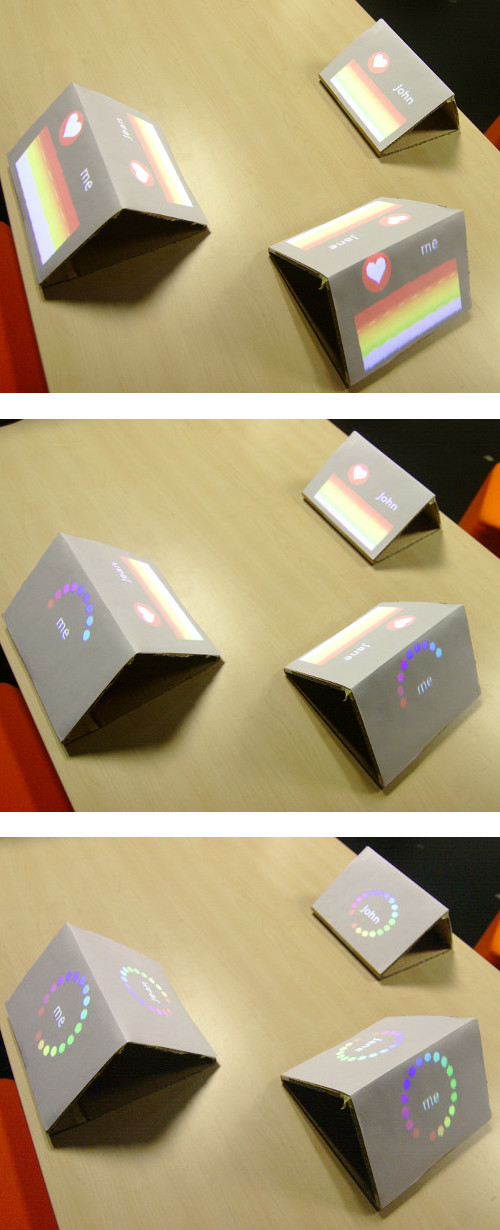}
   \captionof{figure}{Our experimental conditions regarding heart rate (HR) visualizations. \emph{Top}: HR visible by all players. \emph{Middle}: HR visible by the others but not by self. \emph{Bottom}: HR not visible.
\label{fig:conditions}}
\end{minipage} \\
}

To record players' faces, we used a set of 3 Sony PlayStation Eyes
cameras hanging from the ceiling, slightly above the head. These webcams
are cheap -- around 10 dollars the unit at market price -- and yet
provided good images. We used a 640x480 resolution at 30 FPS and a
custom Linux driver that directly fetches raw data (i.e.~the Bayer
matrix) to ensure maximum video quality.

\subsection{Protocol}\label{protocol}

The participants came by 3 to play the card game. After they signed a
consent form and filled a demographic questionnaire, they were taught
the rules of the card game and a ``warm up'' game took place. Once the
players were confident they knew the rules, the SAR system was
switched-on and the meaning of the visualization was explained to them.

Then one of the 3 conditions occurred, participants playing on their
own. After about 10 minutes and a game ended the participants filled a
questionnaire related to social presence (see next section). There were
on average 2 games per condition.

This step was repeated twice, for the 2 other feedback conditions. After
the completion of the 3 conditions, participants filled one last
questionnaire to sense their general feeling about the setup. Overall, a
session lasted approximately 90 minutes.

\subsection{Measures}\label{measures}

Our main metric is composed by the questionnaire given after each
condition occurred, the Social Presence in Gaming Questionnaire (SPGQ)
\citep{Kort2007}. SPGQ is rated on 5-points Likert scales and contains
21 items in total. Its aim is to qualify social presence between players
on three different axis: ``empathy'' (e.g. ``When the others were happy,
I was happy''), ``negative feelings'' (e.g. ``I felt revengeful'') and
``behavioral engagement'' (e.g. ``The others paid close attention to
me'').

Besides those measures, aimed at comparing our experimental conditions,
we also noted participants' reactions while they were playing in order
to gather more insights about what they experienced in regards to the
biofeedback. To do so, the experimenter wrote down the exchanges between
players that mentioned explicitly the displays or their heart activity.

\subsection{Results}\label{results}

We used a Friedman test and post-hoc pairwise Wilcoxon tests adjusted
for multiple comparisons with false discovery rate to compare our 3
heart rate feedback conditions (``HR none'', ``HR others'', ``HR all'').
There was no significant differences in the SPGQ, although we found a
tendency for the ``negative feelings'' score (p $\approx$ 0.1) between
``HR others'' and ``HR all'' conditions. There was more negative
feelings reported in ``HR others'' compared to ``HR all'' condition --
1.19 \emph{vs} 1.06 (SD: 0.87 \emph{vs} 0.64), see Figure
\ref{fig:SPGQ}.

\marginnote{
\begin{minipage}{\marginparwidth}
   \vspace{0ex}
   \includegraphics[width=0.9\marginparwidth]{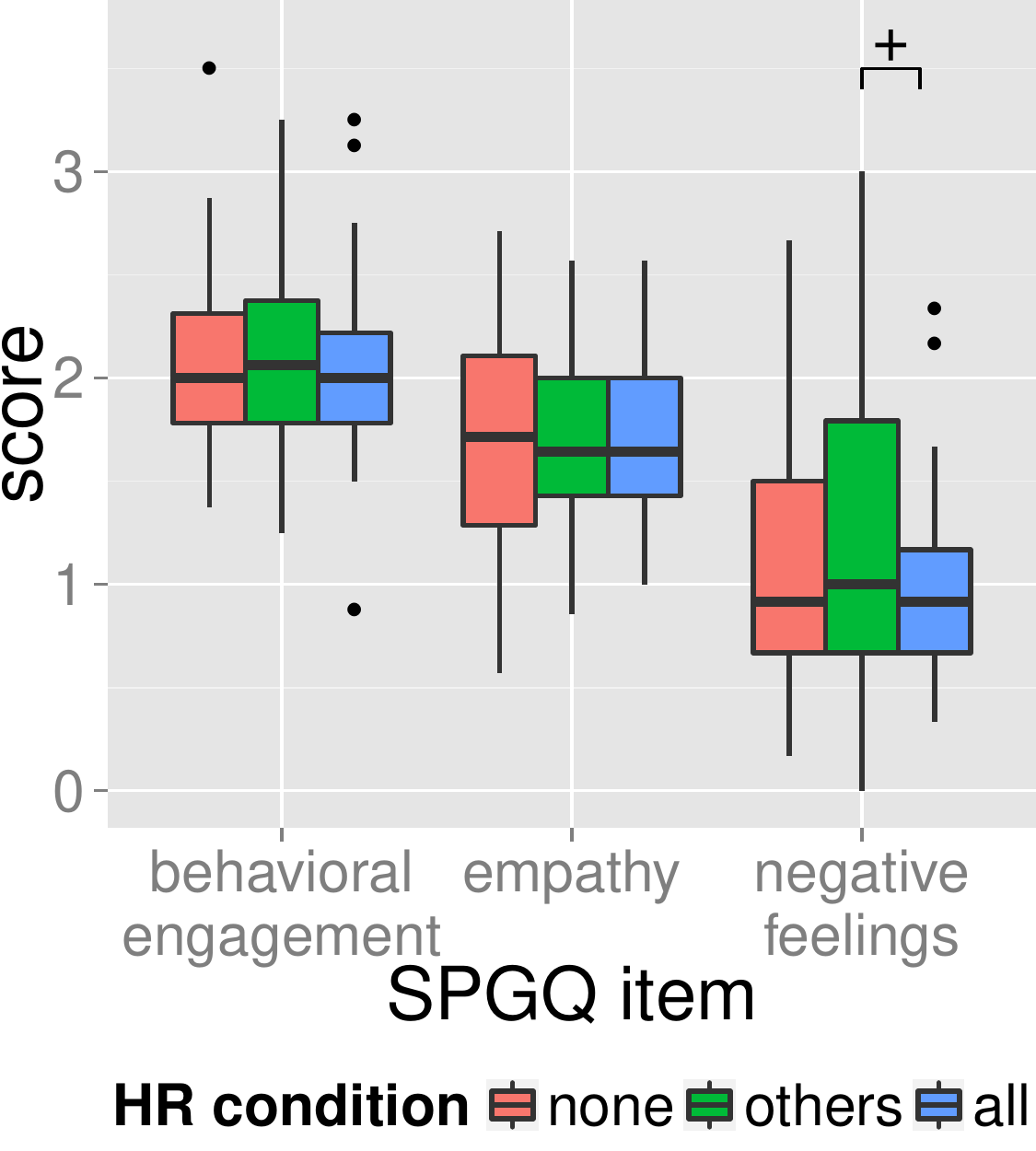}
   \captionof{figure}{Results of the SPGQ questionnaire -- tendency is marked with a "+" sign.
\label{fig:SPGQ}}
\end{minipage}
}

At the end of the pilot study we asked our participants their opinions
about each experimental condition and about the technical aspects of the
setup, using 5-points Likert scales ranging from 0 ``I did not like at
all'' to 4 ``I liked a lot''. The ``HR all'' condition was slightly
favored over the ``HR others'' condition -- 2.83 \emph{vs} 2.78 (SD:
0.79 \emph{vs} 1.06), and the condition with no HR feedback was ranked
last -- mean: 2.44 (SD: 0.86). About the technical aspects of the setup,
participants praised the SAR display -- 3.28 score, SD: 0.89, and were
satisfied with the remote PPG heart rate measure -- 2.89 score, SD:
0.96.

A selection of comments referring to the heart rate feedback that
players said to each other is presented in the side bar page
\pageref{sec:comments}. The proportion of references to emotions or
decision processes, to events in-game or out-game, is representative of
what we annotated during the different games.

As for \emph{how} the physiological feedback was utilized, we observed
two different kinds of players, roughly in equal proportions. First, the
players that did not use explicitly the heart rate display during the
game. Even when they liked to see it, they did not use this information
while interacting with other players. During informal inquiries, those
players reported that the provided heart rate display was hard to
interpret. The second profile is among players that did use the
feedback; participants that knew beforehand the game and the rules were
immediately attracted by the heart rate feedback and remained the most
enthusiast throughout the game.

\section{Discussion}\label{discussion}

Players had a tendency to report more negative feelings in the SPGQ
questionnaire when others could see their heart rate but when themselves
could not. From direct observation, it seems that players used this
asymmetry to ``tease'' themselves, giving false or exaggerated feedback
to the one that could not see by herself or himself the real heart rate.
Sharing the information evenly should prevent social stress -- unless of
course a game designer \emph{wishes} to create a very competitive
gameplay. These findings only partially go along our hypotheses since
players still preferred an asymmetrical feedback rather than no HR
feedback at all when they were asked to rank explicitly the experimental
conditions at the end of the study.

Concerning the utilization of the heart rate display, we did not give
absolute values to avoid a too intrusive feedback, and we used the same
scale for all players in the ``history'' plot to adapt to all
metabolisms. These choices may have prevented some players to comprehend
well the information, but dynamic representation of low-level
physiological signals is still an open question at the moment
\citep{Chanel2015}. Hence, other visualizations should be investigated
as well as other feedback modalities, such as sounds. The profile of
players that used extensively the biofeedback suggests that
physiological signals could be used to add another layer to an existing
game, especially for experts.

Players often associated heart rate to emotions, although in fact this
is \emph{one} among many different traits that could be inferred from
heart rate \citep{Kivikangas2010}. This is easily explained by the
common knowledge surrounding heart activity and by cultural references.
Still, some participants did not hesitate to interpret various -- and
sometimes random -- events in light of the heart rate feedback. During
exchanges between players, most often the firsts to speak were watching
others' signals and wanted to playfully bother their opponents.

\marginpar{%
  \vspace{-80ex} 
    \begin{minipage}{\marginparwidth}

\textbf{Players' comments during the game}
\vspace{1pc}

-- "Your rate is really high now, it's because you're upset!" \\
-- "It's stressful because I don't see my heart!" (\emph{HR others} condition) \\
-- "Look at how his heart's beating, he's going to make a mistake I think..." \\
-- "Damn, I got a huge spike, it's because I won, I killed someone!" \\
-- "You're a bit fast, you look stressed!" \\
-- "I see your plot and I see you bluffing." \\
-- "We're seeing your plot, don't go crazy!" \\
-- "I don't own the game anymore, I cannot bluff..." \\
-- "You saw how it went up suddenly?" / "Yes, it's because I was happy." \\
-- "He said he liked her, and his heart increased..." \\
-- "You're totally busted, I saw it, it increased!" 
\end{minipage}\label{sec:comments}}

\section{Conclusion \& Future works}\label{conclusion-future-works}

We presented a framework that combined remote heart rate sensing and
projection to bring anew an existing board game. We observed that
players did use the physiological feedback over the course of the games,
suggesting that it could improve the richness of the interactions.
Thorough examinations are needed before we could draw solid conclusions
about how the gaming experience is altered; during our study we sensed
how a discrepancy between what is recorded and what is shown to users
could lead to stress, when one's heart rate is seen only by the
\emph{other} players.

The visual feedback was mapped to simple objects. While this projection
was already sufficient to obtain a game room where anybody, at anytime,
could take a seat and start to play, one may venture further into
spatial augmented reality. Heart rate could be projected on board game
elements or on more detailed objects, small avatars for example, e.g.
\citep{Gervais2016}. As for remote sensing, our system can accommodate
many webcams. We were able to incorporate the Microsoft Kinect 2, which
encompasses several persons at once thanks to its wide-angle lens --
ideal for a ``blackjack'' placement.

In the end, game designers can integrate heart rate measurements
directly to the gameplay and develop a new game system. For example, in
a card game a special picture could mask one's heart rate -- it would be
up to the opponents to decide if the player wants to hide
something\ldots{} or just to make them believe so. A common and shared
space, thanks to SAR, could also favor collaboration or competition over
physiological states. The possibilities are limitless, and may be
explored by the players themselves. As matter of fact, not all the
modifications observed by \emph{our} players were related to actual
physiological changes. Some players rightfully reported that the values
changed when the webcam was ``seeing the hair''. Attempting to deceive
opponents is one way to play with physiological signals.

\section{Acknowledgments}\label{acknowledgments}

I thank Jérémy Laviole and Renaud Gervais for their ideas and support
during this project.

\balance{}

\bibliography{biblio_ext}


\begin{thebibliography}{00}


\ifx \showCODEN    \undefined \def \showCODEN     #1{\unskip}     \fi
\ifx \showDOI      \undefined \def \showDOI       #1{{\tt DOI:}\penalty0{#1}\ }
  \fi
\ifx \showISBNx    \undefined \def \showISBNx     #1{\unskip}     \fi
\ifx \showISBNxiii \undefined \def \showISBNxiii  #1{\unskip}     \fi
\ifx \showISSN     \undefined \def \showISSN      #1{\unskip}     \fi
\ifx \showLCCN     \undefined \def \showLCCN      #1{\unskip}     \fi
\ifx \shownote     \undefined \def \shownote      #1{#1}          \fi
\ifx \showarticletitle \undefined \def \showarticletitle #1{#1}   \fi
\ifx \showURL      \undefined \def \showURL       #1{#1}          \fi

\bibitem{Bousefsaf2013}
{Fr\'{e}d\'{e}ric Bousefsaf}, {Choubeila Maaoui}, {and} {Alain Pruski}. 2013.
\newblock \showarticletitle{{Continuous wavelet filtering on webcam
  photoplethysmographic signals to remotely assess the instantaneous heart
  rate}}.
\newblock {\em Biomedical Signal Processing and Control\/} {8}, 6 (Nov. 2013),
  568--574.
\newblock
\showISSN{17468094}


\bibitem{Chanel2015}
{Guillaume Chanel} {and} {Christian M\"{u}hl}. 2015.
\newblock \showarticletitle{{Connecting Brains and Bodies: Applying
  Physiological Computing to Support Social Interaction}}.
\newblock {\em Interacting with Computers\/} (2015).
\newblock
\showISSN{0953-5438}


\bibitem{Dang2010}
{Chi~Tai Dang} {and} {Elisabeth Andr\'{e}}. 2010.
\newblock \showarticletitle{{Surface-poker: multimodality in tabletop games}}.
  In {\em ITS '10}. 251--252.
\newblock
\showISBNx{9781450303996}


\bibitem{Frey2015c}
{J{\'{e}}r{\'{e}}my Frey}. 2015.
\newblock {\em {Leveraging human-computer interactions and social presence with
  physiological computing}}.
\newblock Ph.D. Dissertation. Univ. Bordeaux.
\newblock


\bibitem{Gervais2016}
{Renaud Gervais}, {J{\'{e}}r{\'{e}}my Frey}, {Alexis Gay}, {Fabien Lotte},
  {and} {Martin Hachet}. 2016.
\newblock \showarticletitle{{Tobe: Tangible Out-of-Body Experience}}. In {\em
  TEI '16}.
\newblock


\bibitem{Janssen2011}
{Joris~H Janssen}, {Joyce~H.D.M. Westerink}, {Wijnand~A. IJsselteijn}, {and}
  {Marjolein~D. van~der Zwaag}. 2011.
\newblock \showarticletitle{{The role of physiological computing in
  counteracting loneliness}}. In {\em CHI '11 Workshop: “Brain and Body
  Interfaces: Designing for Meaningful Interaction"}.
\newblock


\bibitem{Kivikangas2010}
{J.~Matias Kivikangas}, {Inger Ekman}, {Guillaume Chanel}, {Simo
  J\"{a}rvel\"{a}}, {Ben Cowley}, {Pentti Henttonen}, {and} {Niklas Ravaja}.
  2010.
\newblock \showarticletitle{{Review on psychophysiological methods in game
  research}}.
\newblock {\em Nordic DiGRA\/} (2010).
\newblock


\bibitem{Kort2007}
{Yvonne A. W.~de Kort}. 2007.
\newblock \showarticletitle{{Digital Games as Social Presence Technology :
  Development of the Social Presence in Gaming Questionnaire ( SPGQ )}}.
\newblock {\em Presence\/} (2007), 1--9.
\newblock
\showISSN{1046-8781}


\bibitem{Laviole2012b}
{J{\'{e}}r{\'{e}}my Laviole} {and} {Martin Hachet}. 2012.
\newblock \showarticletitle{{PapARt: Interactive 3D graphics and multi-touch
  augmented paper for artistic creation}}. In {\em 3DUI '12}. 3--6.
\newblock
\showISBNx{9781467312059}


\bibitem{Lee2014}
{Myungho Lee}, {Kangsoo Kim}, {Hyunghwan Rho}, {and} {Si~Jung Kim}. 2014.
\newblock \showarticletitle{{Empa talk}}. In {\em CHI EA '14}.
\newblock
\showISBNx{9781450324748}


\bibitem{Raskar2001}
{Ramesh Raskar}, {Greg Welch}, {Kok-Lim Low}, {and} {Deepak Bandyopadhyay}.
  2001.
\newblock \showarticletitle{{Shader lamps: Animating real objects with
  image-based illumination}}. In {\em Proceedings of the 12th Eurographics
  Workshop on Rendering Techniques}.
\newblock
\showISBNx{3211837094}


\bibitem{Short1976}
{John Short}, {Bruce Christie}, {and} {Ederyn Williams}. 1976.
\newblock {\em {The Social Psychology of Telecommunications}}.
\newblock Wiley, John \& Sons, Incorporated. 195 pages.
\newblock
\showISBNx{9780471015819}


\bibitem{Slovak2012}
{Petr Slov\'{a}k}, {Joris~H. Janssen}, {and} {Geraldine Fitzpatrick}. 2012.
\newblock \showarticletitle{{Understanding heart rate sharing: towards
  unpacking physiosocial space}}.
\newblock {\em CHI '12\/} (2012), 859--868.
\newblock
\showISBNx{9781450310154}


\bibitem{Walmink2014}
{Wouter Walmink}, {Danielle Wilde}, {and} {Florian~'Floyd' Mueller}. 2014.
\newblock \showarticletitle{{Displaying Heart Rate Data on a Bicycle Helmet to
  Support Social Exertion Experiences}}. In {\em TEI '14}.
\newblock
\showISBNx{9781450326353}


\bibitem{Williams2015}
{Michele~A Williams}, {Asta Roseway}, {Chris O'Dowd}, {Mary Czerwinski}, {and}
  {Meredith~Ringel Morris}. 2015.
\newblock \showarticletitle{{SWARM: An Actuated Wearable for Mediating
  Affect}}. In {\em TEI '15}. 293--300.
\newblock
\showISBNx{9781450333054}


\bibitem{Yamabe2010}
{Tetsuo Yamabe}, {Ilkka Kosunen}, {Inger Ekman}, {Lassi~A. Liikkanen}, {Kai
  Kuikkaniemi}, {and} {Tatsuo Nakajima}. 2010.
\newblock \showarticletitle{{Biofeedback Training with EmoPoker: Controlling
  Emotional Arousal for Better Poker Play}}. In {\em BioS-Play '10}.
\newblock


\end{thebibliography}

\end{document}